\documentclass[aps,prx,showpacs,twocolumn,groupedaddress,superscriptaddress]{revtex4-1}

\usepackage{times,xspace}
\usepackage{graphicx,graphics,color,epsfig}
\usepackage{amsmath}
\usepackage{amssymb}
\usepackage{amsfonts}
\usepackage{amsfonts}
\usepackage{epstopdf}
\usepackage{bm}
\usepackage{hyperref}
\usepackage{color}
\usepackage{float}
\usepackage[normalem]{ulem}
\usepackage{color}

\def\be{\begin{equation}}
\def\ee{\end{equation}}
\def\ba{\begin{eqnarray}}
\def\ea{\end{eqnarray}}

\begin{document}

%\title{Nonthermal excitonic order in photo-doped multiorbital Hubbard systems}
\title{Photoinduced excitonic magnetism in a multiorbital Hubbard system}

\author{Lei Geng}
\affiliation{Department of Physics, University of Fribourg, Fribourg-1700, Switzerland}
\author{Sujay Ray}
%\email{sujay.ray@unifr.ch}
\affiliation{Department of Physics, University of Fribourg, Fribourg-1700, Switzerland}
\author{Philipp Werner}
%\email{philipp.werner@unifr.ch}
\affiliation{Department of Physics, University of Fribourg, Fribourg-1700, Switzerland}

\date{\today}
 
\begin{abstract}
Multiorbital Hubbard models with Hund coupling and crystal-field splitting exhibit an instability toward spin-triplet excitonic order in the parameter regime characterized by strong local spin fluctuations. Upon chemical doping, two distinct types of excitonic ferromagnetism have been reported. Using steady-state nonequilibrium dynamical mean-field theory, we demonstrate that photo-doped half-filled systems can host nonthermal counterparts of these excitonic phases and exhibit a rich phase diagram in the space of photo-doping and crystal field splitting. Photo-doping a spin-triplet excitonic insulator provides a route towards nonequilibrium control of magnetic order. 
\end{abstract}
\vspace{0.5in}

\maketitle

\section{Introduction}

Experiments on various classes of materials have shown that electronic orders, including excitonic order~\cite{Mor2017} and superconductivity~\cite{Fausti2011,Kaiser2014,Mitrano2016,Buzzi2020,Budden2021}, can be transiently enhanced or even induced by photoexcitation. Similar effects have also been demonstrated in theoretical studies of model systems~\cite{Kaneko2019,Ejima2020,Tindall2019,Werner2019,Li2020,Murakami2022,Geng2024}. Photo-doping of Mott insulators has emerged as a particularly promising route to realize nonthermal ordered states, since the Mott gap protects the photo-doped carriers from rapid thermalization~\cite{Murakami2023}. At the same time, these systems are strongly correlated and, even in equilibrium, exhibit rich phase diagrams upon doping~\cite{Dagotto2005,Kunes2015,Hoshino2016}. A large body of work has focused on the realization of $\eta$-pairing states in photo-doped Mott insulators~\cite{Kaneko2019,Werner2019,Li2020,Murakami2022,Ray2023,Ray2024}, but other types of nonthermal orders have also been reported, including magnetic order~\cite{Ray2024}, Kugel-Khomskii order~\cite{Li2018,Ray2024b}, excitonic order~\cite{Werner2020}, chiral superconductivity~\cite{Li2023}, and odd-frequency orbital order~\cite{Werner2021}.

An interesting platform for the study of electronic ordering instabilities is the two-orbital Hubbard model, which provides the minimal description capturing the energy splitting between different local spin states resulting from the Hund coupling $J$. Even in the case of degenerate orbitals, the two-orbital Hubbard model exhibits a rich phase diagram in the space of filling and interaction strength~\cite{Hoshino2015,Hoshino2016}, with antiferromagnetic, ferromagnetic, and orbital-singlet spin-triplet superconducting phases for $J>0$. If a crystal-field splitting $\Delta$ is introduced between the orbitals, the half-filled system exhibits a high-spin/low-spin transition or crossover near $J\sim \Delta$~\cite{Werner2007}. In low-temperature equilibrium states, a spin-triplet type excitonic order appears in the vicinity of this transition~\cite{Kunes2014,Kunes2015,Nasu2016,Hoshino2016}. Away from half-filling, transitions to ferromagnetic excitonic orders have been reported~\cite{Kunes2014b}. 

In this work, we investigate the impact of nonthermal electron populations on the emergence and stability of electronic orders in the half-filled two-orbital Hubbard model. Using nonequilibrium dynamical mean-field theory (DMFT) \cite{Georges1996,Aoki2014}, we demonstrate that nonthermal forms of excitonic order and excitonic magnetism can exist in steady states that mimic a photo-doped system with an effectively cold population of singlon- and triplon-type charge carriers. Our results suggest that photo-doping a spin-triplet excitonic insulator provides a possible route to control magnetism on ultrafast timescales.

The paper is organized as follows: In Sec.~\ref{sec_model}, we introduce the model and the computational method. Section~\ref{sec_results} presents the main results, and Sec.~\ref{sec_conclusions} summarizes our conclusions.

\section{Model and Method}
\label{sec_model}

We consider the two-orbital Hubbard model with Hund coupling and crystal field splitting. The lattice Hamiltonian reads
\begin{align}
&H = -\sum_{\left<ij\right>,\sigma} \sum_{\alpha=1,2} t^\text{hop}_\alpha c^{\dagger}_{i,\alpha\sigma}c_{j,\alpha\sigma} + U\sum_{i}\sum_{\alpha=1,2}n_{i,\alpha\uparrow}n_{i,\alpha\downarrow}  \nonumber \\
&\hspace{0mm} + (U-2J)\sum_{i,\sigma}n_{i,1\sigma}n_{i,2\Bar{\sigma}}  
+ (U-3J)\sum_{i,\sigma}n_{i,1\sigma}n_{i,2\sigma} \nonumber\\
&- \mu \sum_{i}\sum_{\alpha=1,2} n_{i,\alpha} +\frac{\Delta}{2}\sum_{i}(n_{i,2}-n_{i,1}).
\label{eq_1}
\end{align}
Here, $c^{\dagger}_{i,\alpha\sigma}$ ($c_{i,\alpha\sigma}$) creates (annihilates) an electron with spin $\sigma$ in orbital $\alpha$ at site $i$, $n_{i,\alpha\sigma}$ denotes the corresponding spin- and orbital-resolved density, and $n_{i,\alpha}=n_{i,\alpha\uparrow}+n_{i,\alpha\downarrow}$. The orbital-dependent nearest-neighbor hopping amplitude is $t^{\text{hop}}_\alpha$, and we choose $t^{\text{hop}}_2=-t^{\text{hop}}_1$ to realize opposite band dispersions for the two orbitals. The parameter $U$ represents the intra-orbital Hubbard interaction, $J$ the Hund coupling, $\Delta$ the crystal-field splitting, and $\mu$ the chemical potential. For simplicity, we neglect the spin-flip and pair-hopping terms. 

We solve this model using DMFT~\cite{Aoki2014} within the nonequilibrium steady-state (NESS) framework~\cite{Li2021}. In this formalism, each lattice site is weakly coupled to auxiliary fermionic baths, which enables the stabilization of nonthermal distributions of local multiplet states. In the half-filled system, whose equilibrium configuration is dominated by doubly occupied sites, these baths can be used to generate enhanced populations of triply and singly occupied sites (triplons and singlons), effectively mimicking a photo-doped state. To achieve the photo-doping effect, the fermion baths are attached in the energy regions corresponding to the lower and upper Hubbard bands. The density of ``photo-carriers'' is controlled by the bath chemical potentials $\pm\mu_b$, while the bath temperature $T_b$ serves as a tuning parameter for the effective temperature of the triplon and singlon populations~\cite{Li2021,Ray2024}.

DMFT maps the lattice model to a self-consistently determined quantum impurity model with a local term analogous to that of the lattice Hamiltonian and a bath represented by the hybridization function $\Lambda(t,t')$ \cite{Georges1996,Aoki2014}. The simulations are performed for an infinitely connected Bethe lattice, for which the DMFT self-consistency equation simplifies to \cite{Georges1996}
\begin{eqnarray}
    \Lambda(t,t^{\prime})=V G(t,t^{\prime})V + \sum_{b} \Lambda_{b}(t,t^{\prime}) \mathcal{I},\label{eq_selfconsistency}
\label{eq_2}
\end{eqnarray}
with $G(t,t')=-i\langle T_\mathcal{C}\psi(t)\psi^\dagger(t')\rangle$ the local Green's function expressed in terms of the spinor $\psi^\dagger=(c^\dagger_{1\uparrow}, c^\dagger_{2\uparrow}, c^\dagger_{1\downarrow}, c^\dagger_{2\downarrow})$, $V=\text{diag}(v,-v,v,-v)$ and $\mathcal{I}=\text{diag}(1,1,1,1)$. This self-consistency allows us to study uniform (non-staggered) electronic orders. We set the rescaled hopping amplitude $v=|t^\text{hop}|/\sqrt{Z}$ (with $Z\rightarrow \infty$ the coordination number) to unity, so that the noninteracting bandwidth of both bands is $W=4v$. We use $v=1$ as the unit of energy ($\hbar/v\equiv 1/v$ as the unit of time). 

The hybridization function $\Lambda$ has two contributions. The first term on the right hand side of Eq.~\eqref{eq_selfconsistency} represents the effect of the lattice and the second term that of the fermionic baths. The hybridization functions associated with the latter are $-\text{Im}\Lambda^R_b(\omega)=$ $\pi g^2\rho_{b}(\omega)=\Gamma \sqrt{W_{b}^{2}-(\omega - \omega_{b})^{2}}$, with $\Gamma=g^2/W_{b}^{2}$ a dimensionless coupling constant, $\omega_b$ the center of the semi-elliptical bath density of states, and $W_b$ the half-bandwidth of bath $b$. In a steady-state situation, the hybridization functions and Green's functions only depend on the time difference $t-t^{\prime}$. In the half-filled non-crossing approximation (NCA) calculations presented in the main text, we employ two fermionic baths characterized by $\omega_{b} = \pm 5.0$, $W_b = 2.0$, and $g = 0.5$ (corresponding to $\Gamma = 0.0625$). The bath temperature is set to $T_b = 0.02$, and the photo-doping concentration is controlled by varying the bath chemical potential $\mu_b$. In the one-crossing approximation (OCA) calculations, two additional fermionic baths with $\omega_{b} = \pm 20$, $W_b = 5.0$, $g = 0.5$, $\mu_b = 0$, and $T_b = 0.02$ are introduced to suppress non-physical quadruplon and empty states.

%To solve the nonequilibrium DMFT equations, we mainly use the first-order strong-coupling expansion (NCA) \cite{Keiter1971,Grewe1981,Eckstein2010}. This method is numerically efficient and provides a qualitatively correct description in the strongly correlated regime, as we will demonstrate with some benchmarks against second-order (one-crossing approximation, OCA) \cite{Pruschke1989} results obtained with a recently-developed tensor cross interpolation approach \cite{Ritter2024,Kim2025,Geng2025}.  

As in Refs.~\cite{Kunes2015,Werner2020}, we define the spin-triplet excitonic order parameter as
\begin{equation}
\phi^\lambda=\sum_{\sigma\sigma'}\langle c^\dagger_{1\sigma}c_{2\sigma'}\rangle \sigma^\lambda_{\sigma\sigma'}, \label{eq_OP}
\end{equation}
with $\sigma^\lambda_{\sigma\sigma'}$ representing the Pauli matrices for $\lambda=x,y,z$. 
To study excitonic order and possible excitonic magnetism, we apply a small seed field coupling to 
$c^\dagger_{1\uparrow}c_{2\downarrow}-c^\dagger_{1\downarrow}c_{2\uparrow}$ and a small magnetic field to $S_z=n_\uparrow-n_\downarrow$, 
but only during the first few iterations of the DMFT calculation. 
%In addition to the above excitonic order, the self-consistency relation (\ref{eq_selfconsistency}) also allows us to stabilize ferromagnetic order. 
%or staggered spin-triplet order
In order to study antiferromagnetic order, we flip the spin in the self-consistency equation (\ref{eq_selfconsistency}) by replacing $V\rightarrow \tilde V$, with 
\begin{align}
\tilde V &=\left(
\begin{tabular}{cccc}
0 & 0 & $v$ & 0 \\
0 & 0 & 0 & $-v$ \\
$v$ & 0 & 0 & 0 \\
0 & $-v$ & 0 & 0 \\
\end{tabular}
\right).
\label{AFM_V}
\end{align}

To solve the nonequilibrium DMFT equations, we use the first-order strong-coupling expansion (NCA) \cite{Keiter1971,Grewe1981,Eckstein2010}, unless otherwise noted. This method is numerically efficient and provides a qualitatively correct description in the strongly correlated regime, as we will demonstrate with some benchmarks against second-order (OCA) \cite{pruschke1989} results obtained with a recently-developed tensor cross interpolation approach \cite{Ritter2024,Kim2025,Geng2025}.

\begin{figure}[t]
\includegraphics[width=0.45\textwidth]{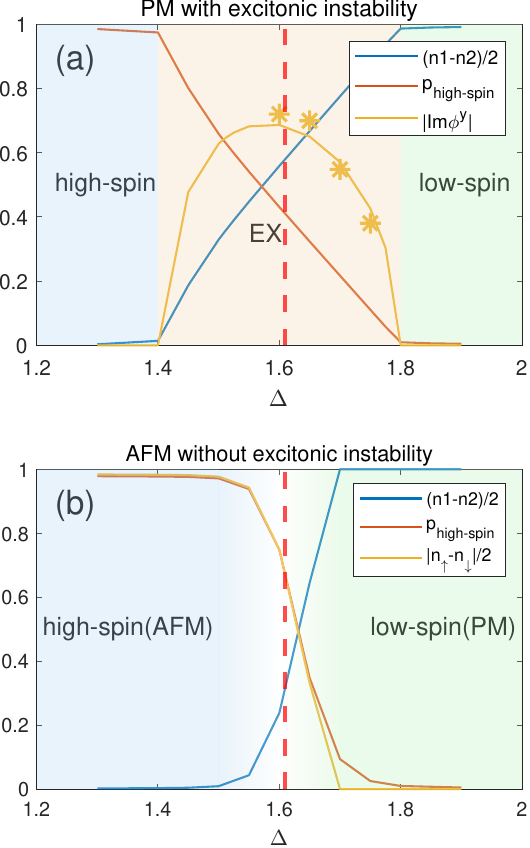}
\caption{
 (a) Order parameter $|\text{Im}\phi^y|$ as a function of $\Delta$ for paramagnetic equilibrium systems at $T_b=0.02$ ($U=14$, $J=0.5$). Also shown are the orbital polarization $(n_1-n_2)/2$ and the probability of high-spin configurations, $p_{\uparrow\uparrow}+p_{\downarrow\downarrow}$. The equilibrium excitonic (EX) phase appears between the high-spin insulating (HS) and low-spin orbitally polarized insulating (LS) phases. The yellow stars indicate $|\text{Im}\phi^y|$ obtained from the OCA calculation. (b) Orbital polarization, high-spin probability, and antiferromagnetic order parameter $|n_{\uparrow}-n_{\downarrow}|/2$ computed for the same parameters as in (a), but without allowing for excitonic order. If both instabilities are considered, the dominant one switches from antiferromagnetic to excitonic order at $\Delta=1.61$ (red dashed line).
}
\label{fig_eq}
\end{figure}

\begin{figure*}
\includegraphics[width=1.0\textwidth]{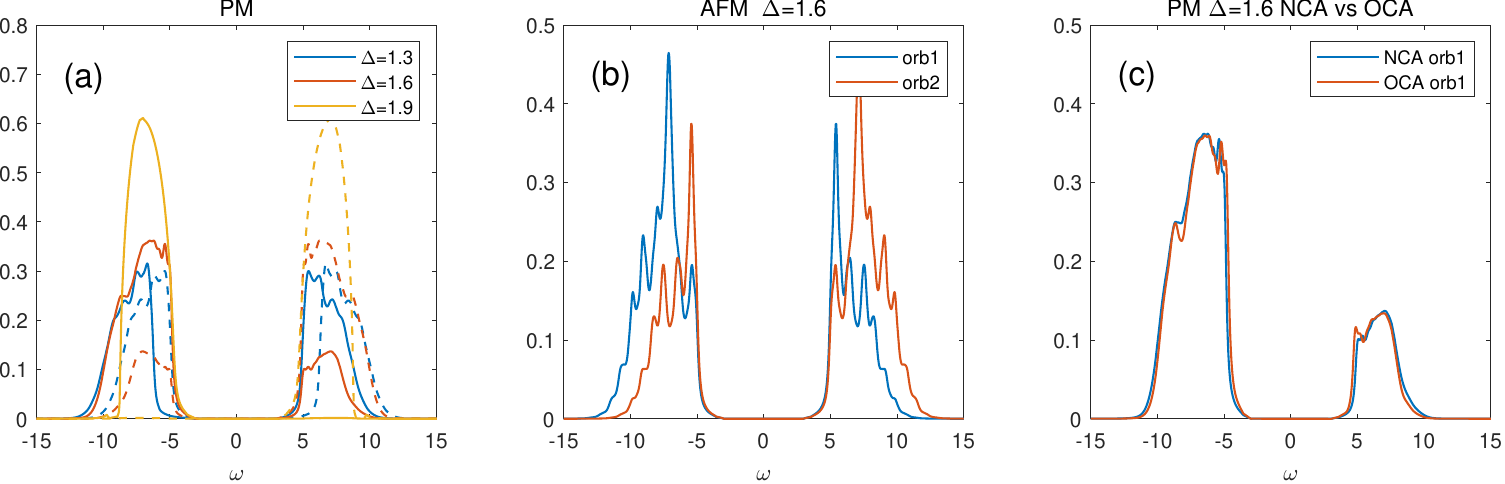}
\caption{
(a) Paramagnetic (PM) equilibrium spectral functions at $T_b=0.02$ for $\Delta=1.3$, 1.6, and 1.9, corresponding to the HS, EX, and LS solutions ($U=14$, $J=0.5$). Solid (dashed) lines represent the spectral functions of orbital 1 (2). (b) Antiferromagnetic spectral functions for the same parameters at $\Delta=1.6$. (c) Comparison of NCA and OCA paramagnetic spectral functions for orbital 1 at $\Delta=1.6$.
}
\label{fig_g_eq}
\end{figure*}

\section{Results}
\label{sec_results}

We consider half-filled strongly correlated systems with $U=14$, $J=0.5$. The photo-doping concentration is measured by the triplon density, which we define as $p_3=\left< n_{1\uparrow}n_{1\downarrow}n_{2}+n_{1}n_{2\uparrow}n_{2\downarrow}-4n_{1\uparrow}n_{1\downarrow}n_{2\uparrow}n_{2\downarrow} \right>$, with $n_{\alpha}=n_{\alpha\uparrow} + n_{\alpha\downarrow}$ the occupation of orbital $\alpha$. 
Since there exists a certain triplon concentration $p_{3,\text{eq}}$ even in the equilibrium state, and both singlons and triplons act as charge carriers in our particle-hole symmetric system, the parameter for the photo-doping concentration is defined as $\delta=2(p_3-p_{3,\text{eq}})$.

\subsection{Equilibrium system}

We start by analyzing the spin-triplet excitonic phase in equilibrium under the paramagnetic constraint, with $\mu_b=0$. Fig.~\ref{fig_eq}(a) shows the $\phi^y$ order parameter, which is purely imaginary for the chosen seed field, as a function of the crystal-field splitting $\Delta$. The excitonic phase emerges in the range $1.4 < \Delta < 1.8$, with the maximum order parameter around $\Delta = 1.6$. In the same panel, we also plot the probability of high-spin configurations, $p_\text{high-spin}=p_{\uparrow\uparrow}+p_{\downarrow\downarrow}$, and the orbital polarization $(n_1-n_2)/2$, which clearly demonstrate that the excitonic (EX) phase appears between a high-spin (HS) insulating state with approximately half-filled orbitals and a low-spin (LS) insulating state with almost complete orbital polarization. Within the excitonic region, $p_\text{high-spin}$ decreases continuously with increasing $\Delta$, while the orbital polarization grows. In the atomic limit, the level crossing between the HS and LS configurations occurs at $\Delta = 3J$, and in the lattice model the transition takes place in close proximity to the same value~\cite{Werner2007}. Thus, the excitonic order develops in the vicinity of the HS-LS transition or crossover in the absence of long-range magnetic order, and reaches its maximum amplitude near this transition.

Our finding of an excitonic order is consistent with earlier results based on numerically exact QMC solvers, which demonstrated that orbital-singlet spin-triplet excitonic order emerges in the HS/LS transition or crossover region of the model~\cite{Kunes2015,Hoshino2016}. It also agrees with a previous NCA-based investigation~\cite{Werner2020}. In particular, Ref.~\cite{Hoshino2016} associated the excitonic order with enhanced local spin fluctuations in this crossover regime. To confirm that NCA captures the correct qualitative behavior, we additionally show the OCA result for the excitonic order parameter at $\Delta \ge 1.6$ by the yellow stars in Fig.~\ref{fig_eq}(a).

In the HS regime, the system develops antiferromagnetic (AFM) order at the simulated temperature $T_b=0.02$. In the absence of excitonic order, this AFM phase persists up to the boundary between the HS and LS states, overlapping with the excitonic phase and thereby leading to a competition between the two instabilities. The AFM tendency of the HS phase is consistent with a previous QMC study~\cite{Hoshino2016}. To analyze this competition, we compute the AFM solution using the setup of Eq.~\eqref{AFM_V}, while suppressing the excitonic instability. The resulting orbital polarization and high-spin probabilities are shown in Fig.~\ref{fig_eq}(b). The AFM order parameter, characterized by the sublattice magnetization $|n_{\uparrow}-n_{\downarrow}|/2$, closely follows the high-spin probability within the AFM phase. Compared to the transition region in Fig.~\ref{fig_eq}(a), the AFM order decays more rapidly with increasing $\Delta$, undergoing a sharp crossover from the HS to the LS regime around $\Delta = 1.5$-$1.7$. In the region where both orders compete, we apply comparable seed fields for the two orders and use the self-consistency with Eq.~\eqref{AFM_V} to identify the dominant one. The results indicate that the transition between the two orders occurs around $\Delta = 1.61$: for smaller $\Delta$, the AFM order dominates, whereas for larger $\Delta$, the excitonic order is realized. This transition point is highlighted by the read dashed lines in Figs.~\ref{fig_eq}(a) and~\ref{fig_eq}(b). We note that although AFM order may prevail in certain regions on the Bethe lattice, magnetic frustration in other lattice geometries could suppress the AFM instability, leaving the excitonic order as the primary instability.

Fig.~\ref{fig_g_eq}(a) shows representative equilibrium spectral functions for the HS ($\Delta=1.3$), EX ($\Delta=1.6$), and LS ($\Delta=1.9$) regimes. These spectra are obtained in the presence of the full and empty fermionic baths at $\omega_b=\pm5$, consistent with the nonequilibrium setup used below ($\mu=\mu_b=0$) \footnote{Without these baths, the excitonic insulator exhibits sharp subband peaks at low temperature.}. The spectral functions for orbital $\alpha=1$ (solid lines) and $\alpha=2$ (dashed lines) are plotted separately. In the HS insulating phase, both orbitals exhibit nearly identical spectra shifted by the level splitting $\Delta$, corresponding to approximately half-filled orbitals. In the excitonic phase, the gap edges for both orbitals become similar, while the Hubbard band width remains comparable to that of the orbitally averaged HS case. Consistent with Fig.~\ref{fig_eq}, the spectra in the excitonic phase reveal a pronounced orbital polarization. In the LS insulator, each orbital exhibits a bandwidth similar to that of a single orbital in the HS phase, but the system now shows almost complete orbital polarization. Notably, the gap size exceeds the width of the Hubbard bands, implying a long lifetime of photo-doped charge carriers~\cite{Sensarma2010,Strohmaier2010,Eckstein2011,Murakami2023}. This long life-time enables the stabilization of cold photo-doped states.

Fig.~\ref{fig_g_eq}(b) displays the spectral function of the AFM solution at $\Delta = 1.6$. The AFM spectrum features a series of sharp spin-polaron peaks. Compared to the peak structure of the paramagnetic excitonic phase in Fig.~\ref{fig_g_eq}(a), the AFM peaks are more widely spaced and not restricted to the edges of the Hubbard bands, but instead span the entire energy range. This behavior is consistent with the single-orbital AFM spectra reported in previous studies~\cite{Taranto2012,Werner2012,Geng2025}.

In Fig.~\ref{fig_g_eq}(c), we compare the NCA and OCA results for the orbital-1 spectral function of the excitonic solution at $\Delta=1.6$. While the NCA spectral function slightly overestimates the gap size compared to OCA and the gap edge features are suppressed, the overall spectral shape in the two approximations is very similar. Since stabilizing some of the excitonic solutions requires a very large number of DMFT iterations~\cite{Kunes2014b}, and multiorbital OCA calculations are computationally much more demanding than NCA calculations, we present NCA results in the remainder of this section.

\subsection{Nonequilibrium system}

\begin{figure}
\includegraphics[width=0.45\textwidth]{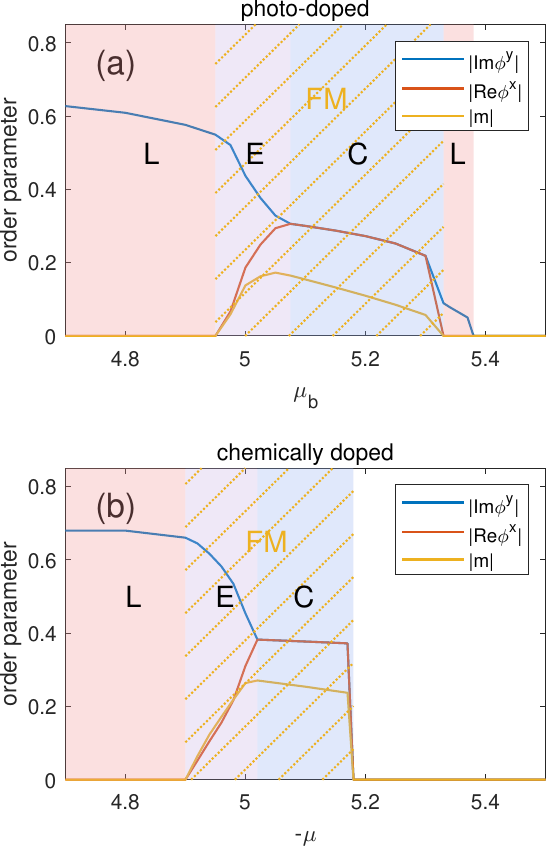}
\caption{
(a) Excitonic order parameters $|\mathrm{Im}\,\phi^y|$, $|\mathrm{Re}\,\phi^x|$, and the magnetization $|m|$ as functions of the bath chemical potential $\mu_b$ for $T_b=0.02$ and $\Delta=1.6$ ($U=14$, $J=0.5$). L, E, and C denote linear, elliptical, and circular polarizations, respectively, following the nomenclature of Ref.~\cite{Kunes2014b}. The yellow hatched region exhibits FM order.
(b) Same quantities calculated as a function of the chemical potential $\mu$ in a chemically doped equilibrium system without baths.
}
\label{fig_photodoped}
\end{figure}

We now turn to the investigation of photo-doped systems, as described by the NESS formalism with bath chemical potentials $\mu_\pm=\pm\mu_b$. The top panel of Fig.~\ref{fig_photodoped}(a) shows the absolute value of $\mathrm{Im}\,\phi^y$ for $\Delta=1.6$, which is close to the optimal crystal-field splitting for the equilibrium excitonic order. (We plot absolute values because in some cases the simulations converged to order parameters with the opposite sign.) Up to $\mu_b=4.95$, the excitonic order remains of the $\phi^y$ type and purely imaginary. Its magnitude is only slightly reduced by the introduction of singlons and triplons, corresponding to at most a few percent photo-doping for these bath chemical potentials. At $\mu_b\simeq4.95$, the $\phi^y$ component of the order parameter suddenly drops, while a nonzero $\phi^x$ component emerges. Simultaneously, a uniform magnetization $m$ develops and rapidly increases with further photo-doping. This state with $|\phi^x|\neq|\phi^y|>0$ persists up to $\mu_b=5.075$, beyond which the magnitudes of $\phi^x$ and $\phi^y$ become identical and the magnetization starts to decrease with increasing $\mu_b$. In the range $\mu_b=5.33$-$5.38$, the system reenters a state with $\mathrm{Im}\,\phi^y\neq0$ and $\phi^x=0$, although the magnitude of the order is significantly smaller than in equilibrium. 

The transitions among these different excitonic insulator phases upon photo-doping are similar to those reported for chemically doped equilibrium systems by Kune\v{s}~\cite{Kunes2014b}. To facilitate the comparison, we analyze our data using an analogous notation. In Ref.~\cite{Kunes2014b}, the order parameters $\phi^-$ and $\phi^+$ were introduced as linear combinations of $\phi^x$ and $\phi^y$. Because $\phi^x$ and $i\phi^y$ are real in our calculations and their signs may occasionally flip between simulations, we define $\phi^\pm$ as $(\phi^+,\phi^-)=(|\phi^x| - |i\phi^y|, |\phi^x| + |i\phi^y|)/2$.
%
%\begin{align}
%\phi^-&=|\phi^x|+|i\phi^y|,\\
%\phi^+&=|\phi^x|-|i\phi^y|.
%\end{align}
%
Using $|\phi^-|$ and $|\phi^+|$, we identify the following photo-doped ordered phases \cite{Kunes2014b}:  
(i) the longitudinal excitonic (L) phase, characterized by $|\phi^-|=|\phi^+|>0$, which corresponds to the photo-doped analogue of the equilibrium excitonic phase;  
(ii) the elliptic excitonic (E) phase, with $|\phi^-|\neq|\phi^+|>0$, found in the range $4.95<\mu_b<5.075$;  
(iii) the circular excitonic (C) phase, where $|\phi^-|>0$ and $|\phi^+|=0$, which appears for $5.075\lesssim\mu_b\lesssim5.33$; and  
(iv) a reentrant L phase for $5.33\lesssim\mu_b\lesssim5.38$.  

Thus, upon photo-doping, the system exhibits the same sequence of excitonic phases as reported by Kune\v{s} for chemical doping \cite{Kunes2014b}, with the exception of the reentrant L phase, which does not appear in the equilibrium phase diagram. Consistent with Ref.~\cite{Kunes2014b}, the E and C phases display ferromagnetic (FM) order and therefore represent excitonic magnetic states. The solutions shown in Fig.~\ref{fig_photodoped}(a) are, however, \textit{nonthermal} steady states stabilized in a photo-doped system with an excess population of singlons and triplons. 

To highlight the close correspondence with chemically doped equilibrium systems, we also plot the equilibrium results as a function of the chemical potential $\mu$ in Fig.~\ref{fig_photodoped}(b). These equilibrium data were obtained using the same NCA solver but without coupling to external baths. The comparison demonstrates that for low (photo)-doping, the three excitonic phases occur %over similar ranges of chemical doping 
%\textcolor{red}{[to show this, we would have to plot the results as function of doping, e. g. in the SM]}
%and 
with comparable order parameter values in both the equilibrium and nonequilibrium systems, while the reentrant L phase only exists in the photo-doped system. % to those reported (for different model choices) in Ref.~\cite{Kunes2014b}. 
Hence, regarding excitonic order and excitonic magnetism, the effect of photo-doping closely mirrors that of chemical doping.  

\begin{figure}
\includegraphics[width=0.47\textwidth]{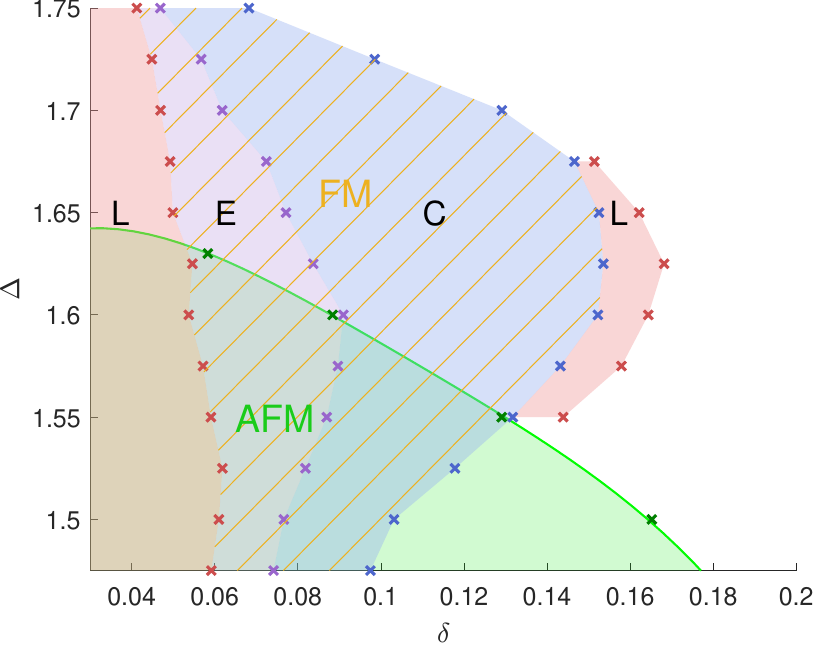}
\caption{Nonequilibrium phase diagram of the photo-doped system in the $\Delta$-$\delta$ plane for $U=14$, $J=0.5$ and $T_b=0.02$.
The red, purple, and blue shaded regions represent the ranges of the L, E, and C phases, respectively, and the yellow hatched region indicates FM order. The green shaded area shows the extent of the AFM phase in a calculation without excitonic orders.
%The red, purple, and blue shaded regions represent the ranges of the L, E, and C phases, respectively, while the green shaded area shows the extent of the AFM phase in a calculation without excitonic orders and the yellow hatched region indicates the FM phase without AFM orders.
}
\label{fig_noneq_delta}
\end{figure}

By varying the value of $\Delta$, we find that the sequence of nonthermal phases L $\rightarrow$ E $\rightarrow$ C persists over nearly the entire range of $\Delta$ where the L-type excitonic phase emerges in %nonequilibrium.
equilibrium. 
The reentrant L-type phase appears only in the vicinity of the maximum excitonic order. The resulting two-dimensional phase diagram in the $\Delta$-$\delta$ plane is shown in Fig.~\ref{fig_noneq_delta}.
We also indicate in the figure the parameter region where AFM order emerges when the excitonic order is suppressed. This magnetic order competes with both the excitonic and excitonic-magnetic orders, and it is the stable phase at small $\Delta$ and small photo-doping (see the red dashed line in Fig.~\ref{fig_eq} for the transition point in the undoped case). 
Depending on the initial value of $\Delta$ and the photo-doping concentration, the photo-excited system can thus undergo transitions from the AFM phase to excitonic-magnetic phases, to the nonmagnetic L-type excitonic phase, or to disordered states.

\begin{figure}
\includegraphics[width=0.45\textwidth]{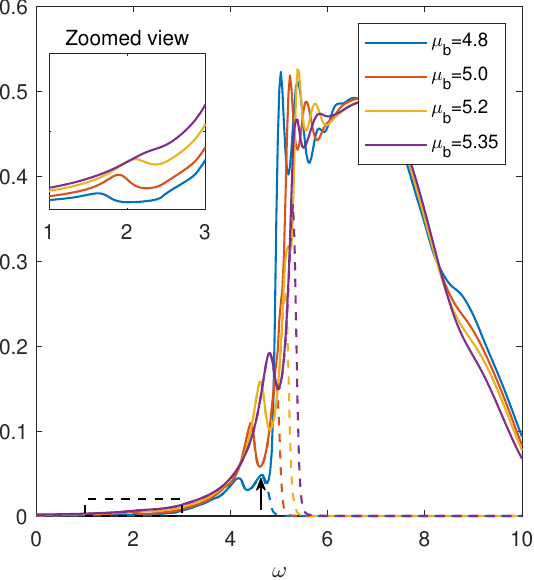}
\caption{Total spectral functions of photo-doped systems for $T_b=0.02$ and $\mu_b=4.8$, 5.0, 5.2, and 5.35 ($\Delta=1.6$, $U=14$, $J=0.5$). The solid line represents the spin- and orbital-summed spectral function $A(\omega)$, while the dashed line shows the corresponding occupation function $A^<(\omega)$. The negative-frequency part is symmetric and corresponds to hole occupation rather than electron occupation. To highlight the small peak around $\omega = 2$, the region marked by the black dashed rectangle is magnified and shown in the upper-left inset.
}
\label{fig_g_photodoped}
\end{figure}

The nonequilibrium spectral functions for four bath chemical potentials corresponding to the L, E, C, and reentrant L phases are shown in Fig.~\ref{fig_g_photodoped}. To facilitate the comparison, we summed over spin and orbital, so that even the ferromagnetic states exhibit spectra symmetric around $\omega=0$. The solid lines show the total spectral functions $A(\omega)$, while the dashed lines indicate the occupied part $A^<(\omega)$~\cite{Aoki2014}. 

Photo-doping generates satellite features inside the gap, which are occupied by triplons (an analogous feature associated with singlons appears for $\omega<0$). These in-gap features correspond to processes in which an electron is removed from a photo-doped triplon state, leaving behind doublon states with various spin and orbital configurations. The positions of the in-gap features can be consistently explained if we assume that the occupied peak near the gap edge (see black arrow in Fig.~\ref{fig_g_photodoped} for the $\mu_b=4.8$ spectrum) corresponds to the electron removal process from the triplon state to the $S=1$ doublon state, $|3\rangle\rightarrow |\!\uparrow,\,\uparrow\rangle$ (or the equivalent process with flipped spins). In the atomic limit, this feature would appear at $\omega=\frac{1}{2}U+\frac{1}{2}J \pm \frac{\Delta}{2}$, where the sign $\pm$ depends on the orbital occupation of the triplon state $|3\rangle$.
The in-gap peak which is shifted lower by about $\Delta\omega=0.5=J$ is then associated with processes of the type $|3\rangle\rightarrow |\!\uparrow,\downarrow\rangle$, since in the atomic limit these have energy $\omega=\frac{1}{2}U-\frac{1}{2}J \pm \frac{\Delta}{2}$. The weak shoulder near $\omega=2$, shown in the inset of Fig.~\ref{fig_g_photodoped}, is instead related to processes of the type $|3\rangle\rightarrow |\uparrow\downarrow,0\rangle$, which in the atomic limit produce features at $\omega=\frac{1}{2}U-\frac{5}{2}J \mp \frac{\Delta}{2}$. For a certain combination of initial and final orbital occupations we thus expect a feature at an energy difference of $\Delta\omega=3J+\Delta=3.1$ relative to the gap edge peak. 

The accumulation of the triplon occupation at the bottom of the band in Fig.~\ref{fig_g_photodoped} suggests effectively cold photo-doped states, whose effective temperatures can be extracted by fitting $A^<(\omega)/A(\omega)$ to a Fermi function. Here, it is important to note that the photo-doped states are nonequilibrium steady states in which the energy released by singlon-triplon recombination processes is balanced by the energy flow into the baths. Hence, we expect that the effective temperatures will be higher than the temperature of the two baths. For the four cases shown in Fig.~\ref{fig_g_photodoped}, the fitted effective inverse temperatures $\beta_{\text{eff}}$ are 39.1, 34.7, 29.7, and 27.6 for $\mu_b = 4.8, 5.0, 5.2,$ and $5.35$, respectively, while the inverse bath temperature is $\beta_{\text{b}} = 50$. These results show that stronger photo-doping induces more heating due to enhanced recombination processes.

\section{Conclusions}
\label{sec_conclusions} 

We employed nonequilibrium dynamical mean-field theory in its steady-state formulation~\cite{Li2020,Li2021} to investigate the influence of photo-doping on the excitonic-insulator phase of the two-orbital Hubbard model. In equilibrium, the excitonic phase of the half-filled system emerges in the spin-transition or crossover region near $\Delta = 3J$ and is characterized by a mixture of high-spin and low-spin doublon states. Within the NESS formalism, photo-doping can be mimicked by coupling the system to electron and hole reservoirs, which induces a nonthermal population of singlon and triplon states. The effective temperature of the resulting steady state can be estimated by fitting a Fermi distribution to the singlon and triplon populations.

We demonstrated that, over a wide range of photo-doping concentrations ($0.05 \lesssim \delta \lesssim 0.15$) and at sufficiently low effective temperatures, excitonic magnetism with nonzero ferromagnetic and excitonic order parameters emerges. Specifically, we identified two distinct types of excitonic magnetism, differentiated by the relative amplitudes of the $\phi^x$ and $\phi^y$ order parameters. These nonthermal steady states closely resemble the equilibrium excitonic magnetic phases previously reported in chemically doped systems \cite{Kunes2014b}. In addition, we identified a reentrant nonmagnetic (L-type) excitonic order, which appears near the optimal value of the crystal field splitting and is unique to the nonequilibrium phase diagram. 

From a technological perspective, the nonthermal magnetic phases are particularly intriguing because the photo-doping concentration can be tuned on ultrafast timescales using laser excitations. In our large-gap system, nonthermal excitonic magnetism persists across nearly the entire range of crystal-field splittings that support excitonic order in equilibrium. Our results thus suggest that spin-triplet excitonic insulators described by strongly correlated multi-orbital Hubbard models constitute a potential platform for nonequilibrium control of magnetic order.

\acknowledgements 
 
The calculations were run on the Beo06 cluster at the University of Fribourg. We acknowledge support from SNSF Grant No. 200021-196966.

\appendix
\section{Phase diagram in the $(\Delta,\mu_b)$ plane}
\label{app_A}
\begin{figure}
\includegraphics[width=0.47\textwidth]{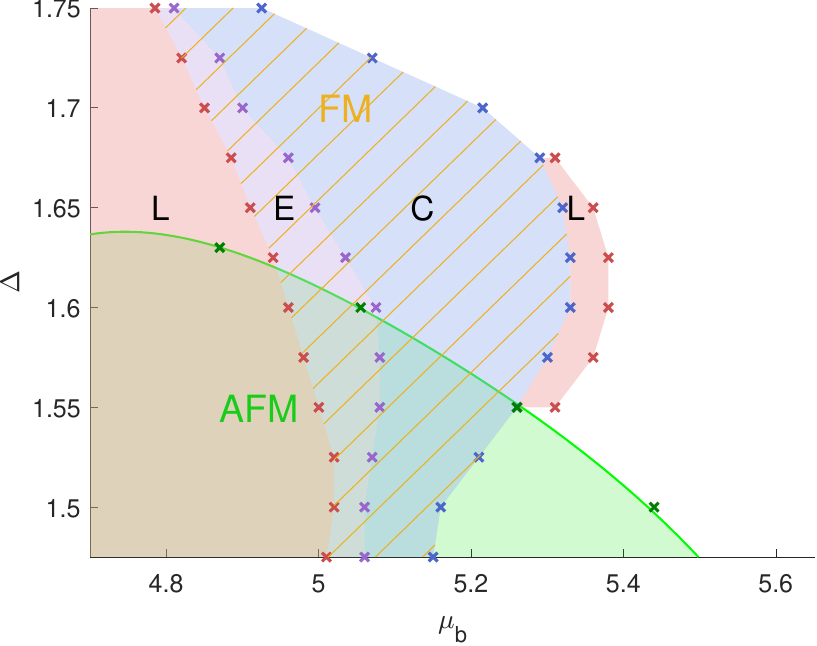}
\caption{Phase diagram of the photo-doped system in the $\Delta$-$\mu_b$ plane for $U=14$, $J=0.5$ and $T_b=0.02$.
The red, purple, and blue shaded regions represent the ranges of the L, E, and C phases, respectively, and the yellow hatched region indicates FM order. The green shaded area shows the extent of the AFM phase in a calculation without excitonic orders.
}
\label{fig_noneq_mu}
\end{figure}
In the main text, we presented the two-dimensional phase diagram in the space of crystal-field splitting $\Delta$ and photo-doping concentration $\delta$, since this representation allows a direct physical interpretation of the results. However, in the actual simulations, the photo-doping concentration is controlled by tuning the bath chemical potential $\mu_b$, which determines the filling of the electron and hole reservoirs in the NESS formalism. For completeness and to facilitate the reproducibility of our results, we also provide in Fig.~\ref{fig_noneq_mu} the phase diagram in the $(\Delta,\mu_b)$ plane. This representation should provide useful reference data for readers who wish to reproduce our data.

\section{Effect of the bath coupling on the equilibrium spectra}
\label{app_B}

\begin{figure}
\includegraphics[width=0.4\textwidth]{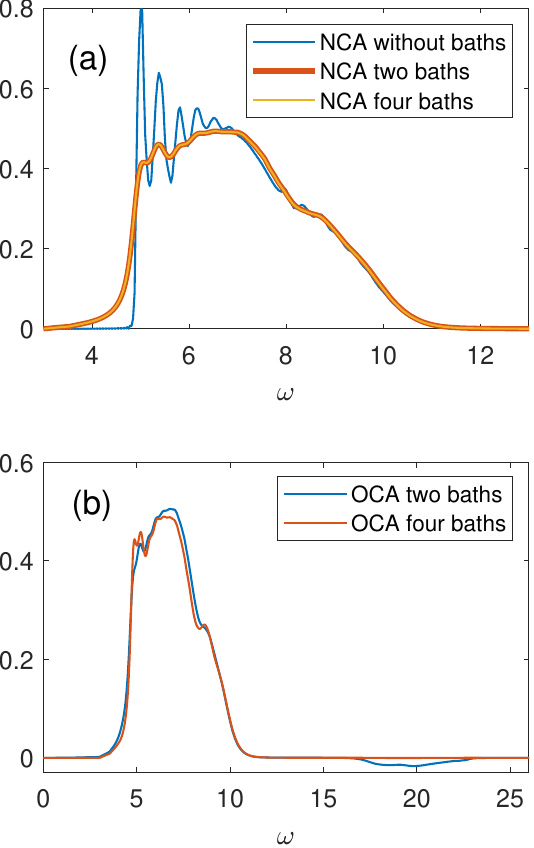}
\caption{(a) Total spectral functions of equilibrium systems calculated by NCA for no-bath, two-bath and four-bath configurations ($\Delta=1.6$, $U=14$, $J=0.5$, $T_b=0.02$). (b) The spectral functions for two-bath and four-bath setups with the same parameters, calculated by OCA.
}
\label{fig_g_bath}
\end{figure}

With the exception of Fig.~\ref{fig_photodoped}(b) (no bath) and the OCA spectrum in Fig.~\ref{fig_g_eq}(c) (two pairs of baths), even for the equilibrium calculations, the system was coupled to one pair of baths (a full and an empty bath), in order to ensure consistency with the photo-doped nonequilibrium configuration. This coupling has a minor effect on the magnitude of the excitonic order parameter, but it can modify the spectral functions by broadening sharp features.

In the OCA calculations, we found that the spectral functions may exhibit small negative regions in the high-energy regime shown in Fig.~\ref{fig_g_bath}(b) if only a single pair of baths is used. To avoid this issue, we coupled the system to two pairs of baths: one pair centered at $\omega_b = \pm 5$ with a bandwidth $W_b = 2$, as described in the main text, and an additional pair centered at $\omega_b = \pm 20$ with a bandwidth $W_b = 5$ (matching the features associated with quadruplon and empty states), with coupling strength $g=0.5$. The latter pair serves primarily to stabilize the self-energy at large high-energy scales and has a negligible effect on the NCA results. 

For reference and reproducibility, we provide in Fig.~\ref{fig_g_bath}(a) the equilibrium NCA spectral functions for the main parameter set ($\Delta = 1.6$, $U = 14$, $J = 0.5$, $T_b = 0.02$), obtained with the following three bath setups: (i) without any bath coupling, (ii) with a single pair of baths at $\omega_b = \pm 5$, $W_b = 2$, and (iii) with two pairs of baths, including an additional high-energy pair at $\omega_b = \pm 20$, $W_b = 5$. The latter two results are almost identical, and they both exhibit a noticeable suppression of the sharp band-edge peaks compared to the uncoupled case. The corresponding excitonic order parameters $|\mathrm{Im}\,\phi^y|$ are 0.6810, 0.6865, and 0.6868 for configurations (i), (ii), and (iii), respectively. These results demonstrate that while the bath coupling modifies the line shape of the spectral function, at the NCA level it has a negligible influence on the magnitude and stability of the excitonic order.

\bibliography{mybibtex}

\begin{thebibliography}{45}%
\makeatletter
\providecommand \@ifxundefined [1]{%
 \@ifx{#1\undefined}
}%
\providecommand \@ifnum [1]{%
 \ifnum #1\expandafter \@firstoftwo
 \else \expandafter \@secondoftwo
 \fi
}%
\providecommand \@ifx [1]{%
 \ifx #1\expandafter \@firstoftwo
 \else \expandafter \@secondoftwo
 \fi
}%
\providecommand \natexlab [1]{#1}%
\providecommand \enquote  [1]{``#1''}%
\providecommand \bibnamefont  [1]{#1}%
\providecommand \bibfnamefont [1]{#1}%
\providecommand \citenamefont [1]{#1}%
\providecommand \href@noop [0]{\@secondoftwo}%
\providecommand \href [0]{\begingroup \@sanitize@url \@href}%
\providecommand \@href[1]{\@@startlink{#1}\@@href}%
\providecommand \@@href[1]{\endgroup#1\@@endlink}%
\providecommand \@sanitize@url [0]{\catcode `\\12\catcode `\$12\catcode `\&12\catcode `\#12\catcode `\^12\catcode `\_12\catcode `\%12\relax}%
\providecommand \@@startlink[1]{}%
\providecommand \@@endlink[0]{}%
\providecommand \url  [0]{\begingroup\@sanitize@url \@url }%
\providecommand \@url [1]{\endgroup\@href {#1}{\urlprefix }}%
\providecommand \urlprefix  [0]{URL }%
\providecommand \Eprint [0]{\href }%
\providecommand \doibase [0]{http://dx.doi.org/}%
\providecommand \selectlanguage [0]{\@gobble}%
\providecommand \bibinfo  [0]{\@secondoftwo}%
\providecommand \bibfield  [0]{\@secondoftwo}%
\providecommand \translation [1]{[#1]}%
\providecommand \BibitemOpen [0]{}%
\providecommand \bibitemStop [0]{}%
\providecommand \bibitemNoStop [0]{.\EOS\space}%
\providecommand \EOS [0]{\spacefactor3000\relax}%
\providecommand \BibitemShut  [1]{\csname bibitem#1\endcsname}%
\let\auto@bib@innerbib\@empty
%</preamble>
\bibitem [{\citenamefont {Mor}\ \emph {et~al.}(2017)\citenamefont {Mor}, \citenamefont {Herzog}, \citenamefont {Gole\ifmmode~\check{z}\else \v{z}\fi{}}, \citenamefont {Werner}, \citenamefont {Eckstein}, \citenamefont {Katayama}, \citenamefont {Nohara}, \citenamefont {Takagi}, \citenamefont {Mizokawa}, \citenamefont {Monney},\ and\ \citenamefont {St\"ahler}}]{Mor2017}%
  \BibitemOpen
  \bibfield  {author} {\bibinfo {author} {\bibfnamefont {S.}~\bibnamefont {Mor}}, \bibinfo {author} {\bibfnamefont {M.}~\bibnamefont {Herzog}}, \bibinfo {author} {\bibfnamefont {D.}~\bibnamefont {Gole\ifmmode~\check{z}\else \v{z}\fi{}}}, \bibinfo {author} {\bibfnamefont {P.}~\bibnamefont {Werner}}, \bibinfo {author} {\bibfnamefont {M.}~\bibnamefont {Eckstein}}, \bibinfo {author} {\bibfnamefont {N.}~\bibnamefont {Katayama}}, \bibinfo {author} {\bibfnamefont {M.}~\bibnamefont {Nohara}}, \bibinfo {author} {\bibfnamefont {H.}~\bibnamefont {Takagi}}, \bibinfo {author} {\bibfnamefont {T.}~\bibnamefont {Mizokawa}}, \bibinfo {author} {\bibfnamefont {C.}~\bibnamefont {Monney}}, \ and\ \bibinfo {author} {\bibfnamefont {J.}~\bibnamefont {St\"ahler}},\ }\href {\doibase 10.1103/PhysRevLett.119.086401} {\bibfield  {journal} {\bibinfo  {journal} {Phys. Rev. Lett.}\ }\textbf {\bibinfo {volume} {119}},\ \bibinfo {pages} {086401} (\bibinfo {year} {2017})}\BibitemShut {NoStop}%
\bibitem [{\citenamefont {Fausti}\ \emph {et~al.}(2011)\citenamefont {Fausti}, \citenamefont {Tobey}, \citenamefont {Dean}, \citenamefont {Kaiser}, \citenamefont {Dienst}, \citenamefont {Hoffmann}, \citenamefont {Pyon}, \citenamefont {Takayama}, \citenamefont {Takagi},\ and\ \citenamefont {Cavalleri}}]{Fausti2011}%
  \BibitemOpen
  \bibfield  {author} {\bibinfo {author} {\bibfnamefont {D.}~\bibnamefont {Fausti}}, \bibinfo {author} {\bibfnamefont {R.~I.}\ \bibnamefont {Tobey}}, \bibinfo {author} {\bibfnamefont {N.}~\bibnamefont {Dean}}, \bibinfo {author} {\bibfnamefont {S.}~\bibnamefont {Kaiser}}, \bibinfo {author} {\bibfnamefont {A.}~\bibnamefont {Dienst}}, \bibinfo {author} {\bibfnamefont {M.~C.}\ \bibnamefont {Hoffmann}}, \bibinfo {author} {\bibfnamefont {S.}~\bibnamefont {Pyon}}, \bibinfo {author} {\bibfnamefont {T.}~\bibnamefont {Takayama}}, \bibinfo {author} {\bibfnamefont {H.}~\bibnamefont {Takagi}}, \ and\ \bibinfo {author} {\bibfnamefont {A.}~\bibnamefont {Cavalleri}},\ }\href {\doibase 10.1126/science.1197294} {\bibfield  {journal} {\bibinfo  {journal} {Science}\ }\textbf {\bibinfo {volume} {331}},\ \bibinfo {pages} {189} (\bibinfo {year} {2011})}\BibitemShut {NoStop}%
\bibitem [{\citenamefont {Kaiser}\ \emph {et~al.}(2014)\citenamefont {Kaiser}, \citenamefont {Hunt}, \citenamefont {Nicoletti}, \citenamefont {Hu}, \citenamefont {Gierz}, \citenamefont {Liu}, \citenamefont {Le~Tacon}, \citenamefont {Loew}, \citenamefont {Haug}, \citenamefont {Keimer},\ and\ \citenamefont {Cavalleri}}]{Kaiser2014}%
  \BibitemOpen
  \bibfield  {author} {\bibinfo {author} {\bibfnamefont {S.}~\bibnamefont {Kaiser}}, \bibinfo {author} {\bibfnamefont {C.~R.}\ \bibnamefont {Hunt}}, \bibinfo {author} {\bibfnamefont {D.}~\bibnamefont {Nicoletti}}, \bibinfo {author} {\bibfnamefont {W.}~\bibnamefont {Hu}}, \bibinfo {author} {\bibfnamefont {I.}~\bibnamefont {Gierz}}, \bibinfo {author} {\bibfnamefont {H.~Y.}\ \bibnamefont {Liu}}, \bibinfo {author} {\bibfnamefont {M.}~\bibnamefont {Le~Tacon}}, \bibinfo {author} {\bibfnamefont {T.}~\bibnamefont {Loew}}, \bibinfo {author} {\bibfnamefont {D.}~\bibnamefont {Haug}}, \bibinfo {author} {\bibfnamefont {B.}~\bibnamefont {Keimer}}, \ and\ \bibinfo {author} {\bibfnamefont {A.}~\bibnamefont {Cavalleri}},\ }\href {\doibase 10.1103/PhysRevB.89.184516} {\bibfield  {journal} {\bibinfo  {journal} {Phys. Rev. B}\ }\textbf {\bibinfo {volume} {89}},\ \bibinfo {pages} {184516} (\bibinfo {year} {2014})}\BibitemShut {NoStop}%
\bibitem [{\citenamefont {Mitrano}\ \emph {et~al.}(2016)\citenamefont {Mitrano}, \citenamefont {Cantaluppi}, \citenamefont {Nicoletti}, \citenamefont {Kaiser}, \citenamefont {Perucchi}, \citenamefont {Lupi}, \citenamefont {Di~Pietro}, \citenamefont {Pontiroli}, \citenamefont {Ricco}, \citenamefont {Clark} \emph {et~al.}}]{Mitrano2016}%
  \BibitemOpen
  \bibfield  {author} {\bibinfo {author} {\bibfnamefont {M.}~\bibnamefont {Mitrano}}, \bibinfo {author} {\bibfnamefont {A.}~\bibnamefont {Cantaluppi}}, \bibinfo {author} {\bibfnamefont {D.}~\bibnamefont {Nicoletti}}, \bibinfo {author} {\bibfnamefont {S.}~\bibnamefont {Kaiser}}, \bibinfo {author} {\bibfnamefont {A.}~\bibnamefont {Perucchi}}, \bibinfo {author} {\bibfnamefont {S.}~\bibnamefont {Lupi}}, \bibinfo {author} {\bibfnamefont {P.}~\bibnamefont {Di~Pietro}}, \bibinfo {author} {\bibfnamefont {D.}~\bibnamefont {Pontiroli}}, \bibinfo {author} {\bibfnamefont {M.}~\bibnamefont {Ricco}}, \bibinfo {author} {\bibfnamefont {S.~R.}\ \bibnamefont {Clark}},  \emph {et~al.},\ }\href {https://www.nature.com/articles/nature16522} {\bibfield  {journal} {\bibinfo  {journal} {Nature}\ }\textbf {\bibinfo {volume} {530}},\ \bibinfo {pages} {461} (\bibinfo {year} {2016})}\BibitemShut {NoStop}%
\bibitem [{\citenamefont {Buzzi}\ \emph {et~al.}(2020)\citenamefont {Buzzi}, \citenamefont {Nicoletti}, \citenamefont {Fechner}, \citenamefont {Tancogne-Dejean}, \citenamefont {Sentef}, \citenamefont {Georges}, \citenamefont {Biesner}, \citenamefont {Uykur}, \citenamefont {Dressel}, \citenamefont {Henderson}, \citenamefont {Siegrist}, \citenamefont {Schlueter}, \citenamefont {Miyagawa}, \citenamefont {Kanoda}, \citenamefont {Nam}, \citenamefont {Ardavan}, \citenamefont {Coulthard}, \citenamefont {Tindall}, \citenamefont {Schlawin}, \citenamefont {Jaksch},\ and\ \citenamefont {Cavalleri}}]{Buzzi2020}%
  \BibitemOpen
  \bibfield  {author} {\bibinfo {author} {\bibfnamefont {M.}~\bibnamefont {Buzzi}}, \bibinfo {author} {\bibfnamefont {D.}~\bibnamefont {Nicoletti}}, \bibinfo {author} {\bibfnamefont {M.}~\bibnamefont {Fechner}}, \bibinfo {author} {\bibfnamefont {N.}~\bibnamefont {Tancogne-Dejean}}, \bibinfo {author} {\bibfnamefont {M.~A.}\ \bibnamefont {Sentef}}, \bibinfo {author} {\bibfnamefont {A.}~\bibnamefont {Georges}}, \bibinfo {author} {\bibfnamefont {T.}~\bibnamefont {Biesner}}, \bibinfo {author} {\bibfnamefont {E.}~\bibnamefont {Uykur}}, \bibinfo {author} {\bibfnamefont {M.}~\bibnamefont {Dressel}}, \bibinfo {author} {\bibfnamefont {A.}~\bibnamefont {Henderson}}, \bibinfo {author} {\bibfnamefont {T.}~\bibnamefont {Siegrist}}, \bibinfo {author} {\bibfnamefont {J.~A.}\ \bibnamefont {Schlueter}}, \bibinfo {author} {\bibfnamefont {K.}~\bibnamefont {Miyagawa}}, \bibinfo {author} {\bibfnamefont {K.}~\bibnamefont {Kanoda}}, \bibinfo {author} {\bibfnamefont {M.-S.}\ \bibnamefont {Nam}}, \bibinfo {author} {\bibfnamefont {A.}~\bibnamefont {Ardavan}}, \bibinfo {author} {\bibfnamefont {J.}~\bibnamefont {Coulthard}}, \bibinfo {author} {\bibfnamefont {J.}~\bibnamefont {Tindall}}, \bibinfo {author} {\bibfnamefont {F.}~\bibnamefont {Schlawin}}, \bibinfo {author} {\bibfnamefont {D.}~\bibnamefont {Jaksch}}, \ and\ \bibinfo {author} {\bibfnamefont {A.}~\bibnamefont {Cavalleri}},\ }\href {\doibase 10.1103/PhysRevX.10.031028} {\bibfield  {journal} {\bibinfo  {journal} {Phys. Rev. X}\ }\textbf {\bibinfo {volume} {10}},\ \bibinfo {pages} {031028} (\bibinfo {year} {2020})}\BibitemShut {NoStop}%
\bibitem [{\citenamefont {Budden}\ \emph {et~al.}(2021)\citenamefont {Budden}, \citenamefont {Gebert}, \citenamefont {Buzzi}, \citenamefont {Jotzu}, \citenamefont {Wang}, \citenamefont {Matsuyama}, \citenamefont {Meier}, \citenamefont {Laplace}, \citenamefont {Pontiroli}, \citenamefont {Ricc{\`o}} \emph {et~al.}}]{Budden2021}%
  \BibitemOpen
  \bibfield  {author} {\bibinfo {author} {\bibfnamefont {M.}~\bibnamefont {Budden}}, \bibinfo {author} {\bibfnamefont {T.}~\bibnamefont {Gebert}}, \bibinfo {author} {\bibfnamefont {M.}~\bibnamefont {Buzzi}}, \bibinfo {author} {\bibfnamefont {G.}~\bibnamefont {Jotzu}}, \bibinfo {author} {\bibfnamefont {E.}~\bibnamefont {Wang}}, \bibinfo {author} {\bibfnamefont {T.}~\bibnamefont {Matsuyama}}, \bibinfo {author} {\bibfnamefont {G.}~\bibnamefont {Meier}}, \bibinfo {author} {\bibfnamefont {Y.}~\bibnamefont {Laplace}}, \bibinfo {author} {\bibfnamefont {D.}~\bibnamefont {Pontiroli}}, \bibinfo {author} {\bibfnamefont {M.}~\bibnamefont {Ricc{\`o}}},  \emph {et~al.},\ }\href {https://www.nature.com/articles/s41567-020-01148-1} {\bibfield  {journal} {\bibinfo  {journal} {Nat. Phys.}\ }\textbf {\bibinfo {volume} {17}},\ \bibinfo {pages} {611} (\bibinfo {year} {2021})}\BibitemShut {NoStop}%
\bibitem [{\citenamefont {Kaneko}\ \emph {et~al.}(2019)\citenamefont {Kaneko}, \citenamefont {Shirakawa}, \citenamefont {Sorella},\ and\ \citenamefont {Yunoki}}]{Kaneko2019}%
  \BibitemOpen
  \bibfield  {author} {\bibinfo {author} {\bibfnamefont {T.}~\bibnamefont {Kaneko}}, \bibinfo {author} {\bibfnamefont {T.}~\bibnamefont {Shirakawa}}, \bibinfo {author} {\bibfnamefont {S.}~\bibnamefont {Sorella}}, \ and\ \bibinfo {author} {\bibfnamefont {S.}~\bibnamefont {Yunoki}},\ }\href {\doibase 10.1103/PhysRevLett.122.077002} {\bibfield  {journal} {\bibinfo  {journal} {Phys. Rev. Lett.}\ }\textbf {\bibinfo {volume} {122}},\ \bibinfo {pages} {077002} (\bibinfo {year} {2019})}\BibitemShut {NoStop}%
\bibitem [{\citenamefont {Ejima}\ \emph {et~al.}(2020)\citenamefont {Ejima}, \citenamefont {Kaneko}, \citenamefont {Lange}, \citenamefont {Yunoki},\ and\ \citenamefont {Fehske}}]{Ejima2020}%
  \BibitemOpen
  \bibfield  {author} {\bibinfo {author} {\bibfnamefont {S.}~\bibnamefont {Ejima}}, \bibinfo {author} {\bibfnamefont {T.}~\bibnamefont {Kaneko}}, \bibinfo {author} {\bibfnamefont {F.}~\bibnamefont {Lange}}, \bibinfo {author} {\bibfnamefont {S.}~\bibnamefont {Yunoki}}, \ and\ \bibinfo {author} {\bibfnamefont {H.}~\bibnamefont {Fehske}},\ }\href {\doibase 10.1103/PhysRevResearch.2.032008} {\bibfield  {journal} {\bibinfo  {journal} {Phys. Rev. Res.}\ }\textbf {\bibinfo {volume} {2}},\ \bibinfo {pages} {032008} (\bibinfo {year} {2020})}\BibitemShut {NoStop}%
\bibitem [{\citenamefont {Tindall}\ \emph {et~al.}(2019)\citenamefont {Tindall}, \citenamefont {Bu\ifmmode~\check{c}\else \v{c}\fi{}a}, \citenamefont {Coulthard},\ and\ \citenamefont {Jaksch}}]{Tindall2019}%
  \BibitemOpen
  \bibfield  {author} {\bibinfo {author} {\bibfnamefont {J.}~\bibnamefont {Tindall}}, \bibinfo {author} {\bibfnamefont {B.}~\bibnamefont {Bu\ifmmode~\check{c}\else \v{c}\fi{}a}}, \bibinfo {author} {\bibfnamefont {J.~R.}\ \bibnamefont {Coulthard}}, \ and\ \bibinfo {author} {\bibfnamefont {D.}~\bibnamefont {Jaksch}},\ }\href {\doibase 10.1103/PhysRevLett.123.030603} {\bibfield  {journal} {\bibinfo  {journal} {Phys. Rev. Lett.}\ }\textbf {\bibinfo {volume} {123}},\ \bibinfo {pages} {030603} (\bibinfo {year} {2019})}\BibitemShut {NoStop}%
\bibitem [{\citenamefont {Werner}\ \emph {et~al.}(2019)\citenamefont {Werner}, \citenamefont {Li}, \citenamefont {Gole\ifmmode~\check{z}\else \v{z}\fi{}},\ and\ \citenamefont {Eckstein}}]{Werner2019}%
  \BibitemOpen
  \bibfield  {author} {\bibinfo {author} {\bibfnamefont {P.}~\bibnamefont {Werner}}, \bibinfo {author} {\bibfnamefont {J.}~\bibnamefont {Li}}, \bibinfo {author} {\bibfnamefont {D.}~\bibnamefont {Gole\ifmmode~\check{z}\else \v{z}\fi{}}}, \ and\ \bibinfo {author} {\bibfnamefont {M.}~\bibnamefont {Eckstein}},\ }\href {\doibase 10.1103/PhysRevB.100.155130} {\bibfield  {journal} {\bibinfo  {journal} {Phys. Rev. B}\ }\textbf {\bibinfo {volume} {100}},\ \bibinfo {pages} {155130} (\bibinfo {year} {2019})}\BibitemShut {NoStop}%
\bibitem [{\citenamefont {Li}\ \emph {et~al.}(2020)\citenamefont {Li}, \citenamefont {Golez}, \citenamefont {Werner},\ and\ \citenamefont {Eckstein}}]{Li2020}%
  \BibitemOpen
  \bibfield  {author} {\bibinfo {author} {\bibfnamefont {J.}~\bibnamefont {Li}}, \bibinfo {author} {\bibfnamefont {D.}~\bibnamefont {Golez}}, \bibinfo {author} {\bibfnamefont {P.}~\bibnamefont {Werner}}, \ and\ \bibinfo {author} {\bibfnamefont {M.}~\bibnamefont {Eckstein}},\ }\href {\doibase 10.1103/PhysRevB.102.165136} {\bibfield  {journal} {\bibinfo  {journal} {Phys. Rev. B}\ }\textbf {\bibinfo {volume} {102}},\ \bibinfo {pages} {165136} (\bibinfo {year} {2020})}\BibitemShut {NoStop}%
\bibitem [{\citenamefont {Murakami}\ \emph {et~al.}(2022)\citenamefont {Murakami}, \citenamefont {Takayoshi}, \citenamefont {Kaneko}, \citenamefont {Sun}, \citenamefont {Gole{\v{z}}}, \citenamefont {Millis},\ and\ \citenamefont {Werner}}]{Murakami2022}%
  \BibitemOpen
  \bibfield  {author} {\bibinfo {author} {\bibfnamefont {Y.}~\bibnamefont {Murakami}}, \bibinfo {author} {\bibfnamefont {S.}~\bibnamefont {Takayoshi}}, \bibinfo {author} {\bibfnamefont {T.}~\bibnamefont {Kaneko}}, \bibinfo {author} {\bibfnamefont {Z.}~\bibnamefont {Sun}}, \bibinfo {author} {\bibfnamefont {D.}~\bibnamefont {Gole{\v{z}}}}, \bibinfo {author} {\bibfnamefont {A.~J.}\ \bibnamefont {Millis}}, \ and\ \bibinfo {author} {\bibfnamefont {P.}~\bibnamefont {Werner}},\ }\href {https://www.nature.com/articles/s42005-021-00799-7} {\bibfield  {journal} {\bibinfo  {journal} {Commun. Phys.}\ }\textbf {\bibinfo {volume} {5}},\ \bibinfo {pages} {23} (\bibinfo {year} {2022})}\BibitemShut {NoStop}%
\bibitem [{\citenamefont {Geng}\ \emph {et~al.}(2024)\citenamefont {Geng}, \citenamefont {Liu}, \citenamefont {Zhang}, \citenamefont {Gole\ifmmode~\check{z}\else \v{z}\fi{}},\ and\ \citenamefont {Peng}}]{Geng2024}%
  \BibitemOpen
  \bibfield  {author} {\bibinfo {author} {\bibfnamefont {L.}~\bibnamefont {Geng}}, \bibinfo {author} {\bibfnamefont {X.}~\bibnamefont {Liu}}, \bibinfo {author} {\bibfnamefont {J.}~\bibnamefont {Zhang}}, \bibinfo {author} {\bibfnamefont {D.}~\bibnamefont {Gole\ifmmode~\check{z}\else \v{z}\fi{}}}, \ and\ \bibinfo {author} {\bibfnamefont {L.-Y.}\ \bibnamefont {Peng}},\ }\href {\doibase 10.1103/PhysRevB.110.115104} {\bibfield  {journal} {\bibinfo  {journal} {Phys. Rev. B}\ }\textbf {\bibinfo {volume} {110}},\ \bibinfo {pages} {115104} (\bibinfo {year} {2024})}\BibitemShut {NoStop}%
\bibitem [{\citenamefont {Murakami}\ \emph {et~al.}(2025)\citenamefont {Murakami}, \citenamefont {Gole\ifmmode~\check{z}\else \v{z}\fi{}}, \citenamefont {Eckstein},\ and\ \citenamefont {Werner}}]{Murakami2023}%
  \BibitemOpen
  \bibfield  {author} {\bibinfo {author} {\bibfnamefont {Y.}~\bibnamefont {Murakami}}, \bibinfo {author} {\bibfnamefont {D.}~\bibnamefont {Gole\ifmmode~\check{z}\else \v{z}\fi{}}}, \bibinfo {author} {\bibfnamefont {M.}~\bibnamefont {Eckstein}}, \ and\ \bibinfo {author} {\bibfnamefont {P.}~\bibnamefont {Werner}},\ }\href {\doibase 10.1103/tkjh-lr83} {\bibfield  {journal} {\bibinfo  {journal} {Rev. Mod. Phys.}\ }\textbf {\bibinfo {volume} {97}},\ \bibinfo {pages} {035001} (\bibinfo {year} {2025})}\BibitemShut {NoStop}%
\bibitem [{\citenamefont {Dagotto}(2005)}]{Dagotto2005}%
  \BibitemOpen
  \bibfield  {author} {\bibinfo {author} {\bibfnamefont {E.}~\bibnamefont {Dagotto}},\ }\href {\doibase 10.1126/science.1107559} {\bibfield  {journal} {\bibinfo  {journal} {Science}\ }\textbf {\bibinfo {volume} {309}},\ \bibinfo {pages} {257} (\bibinfo {year} {2005})}\BibitemShut {NoStop}%
\bibitem [{\citenamefont {Kune{\v{s}}}(2015)}]{Kunes2015}%
  \BibitemOpen
  \bibfield  {author} {\bibinfo {author} {\bibfnamefont {J.}~\bibnamefont {Kune{\v{s}}}},\ }\href {https://iopscience.iop.org/article/10.1088/0953-8984/27/33/333201} {\bibfield  {journal} {\bibinfo  {journal} {J. Phys.: Condens. Matter}\ }\textbf {\bibinfo {volume} {27}},\ \bibinfo {pages} {333201} (\bibinfo {year} {2015})}\BibitemShut {NoStop}%
\bibitem [{\citenamefont {Hoshino}\ and\ \citenamefont {Werner}(2016)}]{Hoshino2016}%
  \BibitemOpen
  \bibfield  {author} {\bibinfo {author} {\bibfnamefont {S.}~\bibnamefont {Hoshino}}\ and\ \bibinfo {author} {\bibfnamefont {P.}~\bibnamefont {Werner}},\ }\href {https://doi.org/10.1103/PhysRevB.93.155161} {\bibfield  {journal} {\bibinfo  {journal} {Phys. Rev. B}\ }\textbf {\bibinfo {volume} {93}},\ \bibinfo {pages} {155161} (\bibinfo {year} {2016})}\BibitemShut {NoStop}%
\bibitem [{\citenamefont {Ray}\ \emph {et~al.}(2023)\citenamefont {Ray}, \citenamefont {Murakami},\ and\ \citenamefont {Werner}}]{Ray2023}%
  \BibitemOpen
  \bibfield  {author} {\bibinfo {author} {\bibfnamefont {S.}~\bibnamefont {Ray}}, \bibinfo {author} {\bibfnamefont {Y.}~\bibnamefont {Murakami}}, \ and\ \bibinfo {author} {\bibfnamefont {P.}~\bibnamefont {Werner}},\ }\href {\doibase 10.1103/PhysRevB.108.174515} {\bibfield  {journal} {\bibinfo  {journal} {Phys. Rev. B}\ }\textbf {\bibinfo {volume} {108}},\ \bibinfo {pages} {174515} (\bibinfo {year} {2023})}\BibitemShut {NoStop}%
\bibitem [{\citenamefont {Ray}\ and\ \citenamefont {Werner}(2024{\natexlab{a}})}]{Ray2024}%
  \BibitemOpen
  \bibfield  {author} {\bibinfo {author} {\bibfnamefont {S.}~\bibnamefont {Ray}}\ and\ \bibinfo {author} {\bibfnamefont {P.}~\bibnamefont {Werner}},\ }\href {\doibase 10.1103/PhysRevB.110.L041109} {\bibfield  {journal} {\bibinfo  {journal} {Phys. Rev. B}\ }\textbf {\bibinfo {volume} {110}},\ \bibinfo {pages} {L041109} (\bibinfo {year} {2024}{\natexlab{a}})}\BibitemShut {NoStop}%
\bibitem [{\citenamefont {Li}\ \emph {et~al.}(2018)\citenamefont {Li}, \citenamefont {Strand}, \citenamefont {Werner},\ and\ \citenamefont {Eckstein}}]{Li2018}%
  \BibitemOpen
  \bibfield  {author} {\bibinfo {author} {\bibfnamefont {J.}~\bibnamefont {Li}}, \bibinfo {author} {\bibfnamefont {H.~U.}\ \bibnamefont {Strand}}, \bibinfo {author} {\bibfnamefont {P.}~\bibnamefont {Werner}}, \ and\ \bibinfo {author} {\bibfnamefont {M.}~\bibnamefont {Eckstein}},\ }\href {https://www.nature.com/articles/s41467-018-07051-x} {\bibfield  {journal} {\bibinfo  {journal} {Nat. Commun.}\ }\textbf {\bibinfo {volume} {9}},\ \bibinfo {pages} {4581} (\bibinfo {year} {2018})}\BibitemShut {NoStop}%
\bibitem [{\citenamefont {Ray}\ and\ \citenamefont {Werner}(2024{\natexlab{b}})}]{Ray2024b}%
  \BibitemOpen
  \bibfield  {author} {\bibinfo {author} {\bibfnamefont {S.}~\bibnamefont {Ray}}\ and\ \bibinfo {author} {\bibfnamefont {P.}~\bibnamefont {Werner}},\ }\href {\doibase 10.1103/PhysRevB.110.214433} {\bibfield  {journal} {\bibinfo  {journal} {Phys. Rev. B}\ }\textbf {\bibinfo {volume} {110}},\ \bibinfo {pages} {214433} (\bibinfo {year} {2024}{\natexlab{b}})}\BibitemShut {NoStop}%
\bibitem [{\citenamefont {Werner}\ and\ \citenamefont {Murakami}(2020)}]{Werner2020}%
  \BibitemOpen
  \bibfield  {author} {\bibinfo {author} {\bibfnamefont {P.}~\bibnamefont {Werner}}\ and\ \bibinfo {author} {\bibfnamefont {Y.}~\bibnamefont {Murakami}},\ }\href {\doibase 10.1103/PhysRevB.102.241103} {\bibfield  {journal} {\bibinfo  {journal} {Phys. Rev. B}\ }\textbf {\bibinfo {volume} {102}},\ \bibinfo {pages} {241103} (\bibinfo {year} {2020})}\BibitemShut {NoStop}%
\bibitem [{\citenamefont {Li}\ \emph {et~al.}(2023)\citenamefont {Li}, \citenamefont {M\"uller}, \citenamefont {Kim}, \citenamefont {L\"auchli},\ and\ \citenamefont {Werner}}]{Li2023}%
  \BibitemOpen
  \bibfield  {author} {\bibinfo {author} {\bibfnamefont {J.}~\bibnamefont {Li}}, \bibinfo {author} {\bibfnamefont {M.}~\bibnamefont {M\"uller}}, \bibinfo {author} {\bibfnamefont {A.~J.}\ \bibnamefont {Kim}}, \bibinfo {author} {\bibfnamefont {A.~M.}\ \bibnamefont {L\"auchli}}, \ and\ \bibinfo {author} {\bibfnamefont {P.}~\bibnamefont {Werner}},\ }\href {\doibase 10.1103/PhysRevB.107.205115} {\bibfield  {journal} {\bibinfo  {journal} {Phys. Rev. B}\ }\textbf {\bibinfo {volume} {107}},\ \bibinfo {pages} {205115} (\bibinfo {year} {2023})}\BibitemShut {NoStop}%
\bibitem [{\citenamefont {Werner}\ and\ \citenamefont {Murakami}(2021)}]{Werner2021}%
  \BibitemOpen
  \bibfield  {author} {\bibinfo {author} {\bibfnamefont {P.}~\bibnamefont {Werner}}\ and\ \bibinfo {author} {\bibfnamefont {Y.}~\bibnamefont {Murakami}},\ }\href {\doibase 10.1103/PhysRevB.104.L201101} {\bibfield  {journal} {\bibinfo  {journal} {Phys. Rev. B}\ }\textbf {\bibinfo {volume} {104}},\ \bibinfo {pages} {L201101} (\bibinfo {year} {2021})}\BibitemShut {NoStop}%
\bibitem [{\citenamefont {Hoshino}\ and\ \citenamefont {Werner}(2015)}]{Hoshino2015}%
  \BibitemOpen
  \bibfield  {author} {\bibinfo {author} {\bibfnamefont {S.}~\bibnamefont {Hoshino}}\ and\ \bibinfo {author} {\bibfnamefont {P.}~\bibnamefont {Werner}},\ }\href {\doibase 10.1103/PhysRevLett.115.247001} {\bibfield  {journal} {\bibinfo  {journal} {Phys. Rev. Lett.}\ }\textbf {\bibinfo {volume} {115}},\ \bibinfo {pages} {247001} (\bibinfo {year} {2015})}\BibitemShut {NoStop}%
\bibitem [{\citenamefont {Werner}\ and\ \citenamefont {Millis}(2007)}]{Werner2007}%
  \BibitemOpen
  \bibfield  {author} {\bibinfo {author} {\bibfnamefont {P.}~\bibnamefont {Werner}}\ and\ \bibinfo {author} {\bibfnamefont {A.~J.}\ \bibnamefont {Millis}},\ }\href {\doibase 10.1103/PhysRevLett.99.126405} {\bibfield  {journal} {\bibinfo  {journal} {Phys. Rev. Lett.}\ }\textbf {\bibinfo {volume} {99}},\ \bibinfo {pages} {126405} (\bibinfo {year} {2007})}\BibitemShut {NoStop}%
\bibitem [{\citenamefont {Kune\ifmmode~\check{s}\else \v{s}\fi{}}\ and\ \citenamefont {Augustinsk\'y}(2014)}]{Kunes2014}%
  \BibitemOpen
  \bibfield  {author} {\bibinfo {author} {\bibfnamefont {J.}~\bibnamefont {Kune\ifmmode~\check{s}\else \v{s}\fi{}}}\ and\ \bibinfo {author} {\bibfnamefont {P.}~\bibnamefont {Augustinsk\'y}},\ }\href {\doibase 10.1103/PhysRevB.89.115134} {\bibfield  {journal} {\bibinfo  {journal} {Phys. Rev. B}\ }\textbf {\bibinfo {volume} {89}},\ \bibinfo {pages} {115134} (\bibinfo {year} {2014})}\BibitemShut {NoStop}%
\bibitem [{\citenamefont {Nasu}\ \emph {et~al.}(2016)\citenamefont {Nasu}, \citenamefont {Watanabe}, \citenamefont {Naka},\ and\ \citenamefont {Ishihara}}]{Nasu2016}%
  \BibitemOpen
  \bibfield  {author} {\bibinfo {author} {\bibfnamefont {J.}~\bibnamefont {Nasu}}, \bibinfo {author} {\bibfnamefont {T.}~\bibnamefont {Watanabe}}, \bibinfo {author} {\bibfnamefont {M.}~\bibnamefont {Naka}}, \ and\ \bibinfo {author} {\bibfnamefont {S.}~\bibnamefont {Ishihara}},\ }\href {\doibase 10.1103/PhysRevB.93.205136} {\bibfield  {journal} {\bibinfo  {journal} {Phys. Rev. B}\ }\textbf {\bibinfo {volume} {93}},\ \bibinfo {pages} {205136} (\bibinfo {year} {2016})}\BibitemShut {NoStop}%
\bibitem [{\citenamefont {Kune\ifmmode~\check{s}\else \v{s}\fi{}}(2014)}]{Kunes2014b}%
  \BibitemOpen
  \bibfield  {author} {\bibinfo {author} {\bibfnamefont {J.}~\bibnamefont {Kune\ifmmode~\check{s}\else \v{s}\fi{}}},\ }\href {\doibase 10.1103/PhysRevB.90.235140} {\bibfield  {journal} {\bibinfo  {journal} {Phys. Rev. B}\ }\textbf {\bibinfo {volume} {90}},\ \bibinfo {pages} {235140} (\bibinfo {year} {2014})}\BibitemShut {NoStop}%
\bibitem [{\citenamefont {Georges}\ \emph {et~al.}(1996)\citenamefont {Georges}, \citenamefont {Kotliar}, \citenamefont {Krauth},\ and\ \citenamefont {Rozenberg}}]{Georges1996}%
  \BibitemOpen
  \bibfield  {author} {\bibinfo {author} {\bibfnamefont {A.}~\bibnamefont {Georges}}, \bibinfo {author} {\bibfnamefont {G.}~\bibnamefont {Kotliar}}, \bibinfo {author} {\bibfnamefont {W.}~\bibnamefont {Krauth}}, \ and\ \bibinfo {author} {\bibfnamefont {M.~J.}\ \bibnamefont {Rozenberg}},\ }\href {\doibase 10.1103/RevModPhys.68.13} {\bibfield  {journal} {\bibinfo  {journal} {Rev. Mod. Phys.}\ }\textbf {\bibinfo {volume} {68}},\ \bibinfo {pages} {13} (\bibinfo {year} {1996})}\BibitemShut {NoStop}%
\bibitem [{\citenamefont {Aoki}\ \emph {et~al.}(2014)\citenamefont {Aoki}, \citenamefont {Tsuji}, \citenamefont {Eckstein}, \citenamefont {Kollar}, \citenamefont {Oka},\ and\ \citenamefont {Werner}}]{Aoki2014}%
  \BibitemOpen
  \bibfield  {author} {\bibinfo {author} {\bibfnamefont {H.}~\bibnamefont {Aoki}}, \bibinfo {author} {\bibfnamefont {N.}~\bibnamefont {Tsuji}}, \bibinfo {author} {\bibfnamefont {M.}~\bibnamefont {Eckstein}}, \bibinfo {author} {\bibfnamefont {M.}~\bibnamefont {Kollar}}, \bibinfo {author} {\bibfnamefont {T.}~\bibnamefont {Oka}}, \ and\ \bibinfo {author} {\bibfnamefont {P.}~\bibnamefont {Werner}},\ }\href {\doibase 10.1103/RevModPhys.86.779} {\bibfield  {journal} {\bibinfo  {journal} {Rev. Mod. Phys.}\ }\textbf {\bibinfo {volume} {86}},\ \bibinfo {pages} {779} (\bibinfo {year} {2014})}\BibitemShut {NoStop}%
\bibitem [{\citenamefont {Li}\ and\ \citenamefont {Eckstein}(2021)}]{Li2021}%
  \BibitemOpen
  \bibfield  {author} {\bibinfo {author} {\bibfnamefont {J.}~\bibnamefont {Li}}\ and\ \bibinfo {author} {\bibfnamefont {M.}~\bibnamefont {Eckstein}},\ }\href {\doibase 10.1103/PhysRevB.103.045133} {\bibfield  {journal} {\bibinfo  {journal} {Phys. Rev. B}\ }\textbf {\bibinfo {volume} {103}},\ \bibinfo {pages} {045133} (\bibinfo {year} {2021})}\BibitemShut {NoStop}%
\bibitem [{\citenamefont {Keiter}\ and\ \citenamefont {Kimball}(1971)}]{Keiter1971}%
  \BibitemOpen
  \bibfield  {author} {\bibinfo {author} {\bibfnamefont {H.}~\bibnamefont {Keiter}}\ and\ \bibinfo {author} {\bibfnamefont {J.}~\bibnamefont {Kimball}},\ }\href {https://doi.org/10.1063/1.1660293} {\bibfield  {journal} {\bibinfo  {journal} {J. Appl. Phys.}\ }\textbf {\bibinfo {volume} {42}},\ \bibinfo {pages} {1460} (\bibinfo {year} {1971})}\BibitemShut {NoStop}%
\bibitem [{\citenamefont {Grewe}\ and\ \citenamefont {Keiter}(1981)}]{Grewe1981}%
  \BibitemOpen
  \bibfield  {author} {\bibinfo {author} {\bibfnamefont {N.}~\bibnamefont {Grewe}}\ and\ \bibinfo {author} {\bibfnamefont {H.}~\bibnamefont {Keiter}},\ }\href {https://doi.org/10.1103/PhysRevB.24.4420} {\bibfield  {journal} {\bibinfo  {journal} {Phys. Rev. B}\ }\textbf {\bibinfo {volume} {24}},\ \bibinfo {pages} {4420} (\bibinfo {year} {1981})}\BibitemShut {NoStop}%
\bibitem [{\citenamefont {Eckstein}\ and\ \citenamefont {Werner}(2010)}]{Eckstein2010}%
  \BibitemOpen
  \bibfield  {author} {\bibinfo {author} {\bibfnamefont {M.}~\bibnamefont {Eckstein}}\ and\ \bibinfo {author} {\bibfnamefont {P.}~\bibnamefont {Werner}},\ }\href {\doibase 10.1103/PhysRevB.82.115115} {\bibfield  {journal} {\bibinfo  {journal} {Phys. Rev. B}\ }\textbf {\bibinfo {volume} {82}},\ \bibinfo {pages} {115115} (\bibinfo {year} {2010})}\BibitemShut {NoStop}%
\bibitem [{\citenamefont {Pruschke}\ and\ \citenamefont {Grewe}(1989)}]{pruschke1989}%
  \BibitemOpen
  \bibfield  {author} {\bibinfo {author} {\bibfnamefont {T.}~\bibnamefont {Pruschke}}\ and\ \bibinfo {author} {\bibfnamefont {N.}~\bibnamefont {Grewe}},\ }\href {https://link.springer.com/article/10.1007/BF01311391} {\bibfield  {journal} {\bibinfo  {journal} {Z. Phys. B}\ }\textbf {\bibinfo {volume} {74}},\ \bibinfo {pages} {439} (\bibinfo {year} {1989})}\BibitemShut {NoStop}%
\bibitem [{\citenamefont {Ritter}\ \emph {et~al.}(2024)\citenamefont {Ritter}, \citenamefont {N\'u\~nez Fern\'andez}, \citenamefont {Wallerberger}, \citenamefont {von Delft}, \citenamefont {Shinaoka},\ and\ \citenamefont {Waintal}}]{Ritter2024}%
  \BibitemOpen
  \bibfield  {author} {\bibinfo {author} {\bibfnamefont {M.~K.}\ \bibnamefont {Ritter}}, \bibinfo {author} {\bibfnamefont {Y.}~\bibnamefont {N\'u\~nez Fern\'andez}}, \bibinfo {author} {\bibfnamefont {M.}~\bibnamefont {Wallerberger}}, \bibinfo {author} {\bibfnamefont {J.}~\bibnamefont {von Delft}}, \bibinfo {author} {\bibfnamefont {H.}~\bibnamefont {Shinaoka}}, \ and\ \bibinfo {author} {\bibfnamefont {X.}~\bibnamefont {Waintal}},\ }\href {\doibase 10.1103/PhysRevLett.132.056501} {\bibfield  {journal} {\bibinfo  {journal} {Phys. Rev. Lett.}\ }\textbf {\bibinfo {volume} {132}},\ \bibinfo {pages} {056501} (\bibinfo {year} {2024})}\BibitemShut {NoStop}%
\bibitem [{\citenamefont {Kim}\ and\ \citenamefont {Werner}(2025)}]{Kim2025}%
  \BibitemOpen
  \bibfield  {author} {\bibinfo {author} {\bibfnamefont {A.~J.}\ \bibnamefont {Kim}}\ and\ \bibinfo {author} {\bibfnamefont {P.}~\bibnamefont {Werner}},\ }\href {\doibase 10.1103/PhysRevB.111.125120} {\bibfield  {journal} {\bibinfo  {journal} {Phys. Rev. B}\ }\textbf {\bibinfo {volume} {111}},\ \bibinfo {pages} {125120} (\bibinfo {year} {2025})}\BibitemShut {NoStop}%
\bibitem [{\citenamefont {Geng}\ \emph {et~al.}(2025)\citenamefont {Geng}, \citenamefont {Kim},\ and\ \citenamefont {Werner}}]{Geng2025}%
  \BibitemOpen
  \bibfield  {author} {\bibinfo {author} {\bibfnamefont {L.}~\bibnamefont {Geng}}, \bibinfo {author} {\bibfnamefont {A.~J.}\ \bibnamefont {Kim}}, \ and\ \bibinfo {author} {\bibfnamefont {P.}~\bibnamefont {Werner}},\ }\href {https://doi.org/10.48550/arXiv.2507.20385} {\bibfield  {journal} {\bibinfo  {journal} {arXiv preprint arXiv:2507.20385}\ } (\bibinfo {year} {2025})}\BibitemShut {NoStop}%
\bibitem [{Note1()}]{Note1}%
  \BibitemOpen
  \bibinfo {note} {Without these baths, the excitonic insulator exhibits sharp subband peaks at low temperature.}\BibitemShut {Stop}%
\bibitem [{\citenamefont {Sensarma}\ \emph {et~al.}(2010)\citenamefont {Sensarma}, \citenamefont {Pekker}, \citenamefont {Altman}, \citenamefont {Demler}, \citenamefont {Strohmaier}, \citenamefont {Greif}, \citenamefont {J\"ordens}, \citenamefont {Tarruell}, \citenamefont {Moritz},\ and\ \citenamefont {Esslinger}}]{Sensarma2010}%
  \BibitemOpen
  \bibfield  {author} {\bibinfo {author} {\bibfnamefont {R.}~\bibnamefont {Sensarma}}, \bibinfo {author} {\bibfnamefont {D.}~\bibnamefont {Pekker}}, \bibinfo {author} {\bibfnamefont {E.}~\bibnamefont {Altman}}, \bibinfo {author} {\bibfnamefont {E.}~\bibnamefont {Demler}}, \bibinfo {author} {\bibfnamefont {N.}~\bibnamefont {Strohmaier}}, \bibinfo {author} {\bibfnamefont {D.}~\bibnamefont {Greif}}, \bibinfo {author} {\bibfnamefont {R.}~\bibnamefont {J\"ordens}}, \bibinfo {author} {\bibfnamefont {L.}~\bibnamefont {Tarruell}}, \bibinfo {author} {\bibfnamefont {H.}~\bibnamefont {Moritz}}, \ and\ \bibinfo {author} {\bibfnamefont {T.}~\bibnamefont {Esslinger}},\ }\href {\doibase 10.1103/PhysRevB.82.224302} {\bibfield  {journal} {\bibinfo  {journal} {Phys. Rev. B}\ }\textbf {\bibinfo {volume} {82}},\ \bibinfo {pages} {224302} (\bibinfo {year} {2010})}\BibitemShut {NoStop}%
\bibitem [{\citenamefont {Strohmaier}\ \emph {et~al.}(2010)\citenamefont {Strohmaier}, \citenamefont {Greif}, \citenamefont {J\"ordens}, \citenamefont {Tarruell}, \citenamefont {Moritz}, \citenamefont {Esslinger}, \citenamefont {Sensarma}, \citenamefont {Pekker}, \citenamefont {Altman},\ and\ \citenamefont {Demler}}]{Strohmaier2010}%
  \BibitemOpen
  \bibfield  {author} {\bibinfo {author} {\bibfnamefont {N.}~\bibnamefont {Strohmaier}}, \bibinfo {author} {\bibfnamefont {D.}~\bibnamefont {Greif}}, \bibinfo {author} {\bibfnamefont {R.}~\bibnamefont {J\"ordens}}, \bibinfo {author} {\bibfnamefont {L.}~\bibnamefont {Tarruell}}, \bibinfo {author} {\bibfnamefont {H.}~\bibnamefont {Moritz}}, \bibinfo {author} {\bibfnamefont {T.}~\bibnamefont {Esslinger}}, \bibinfo {author} {\bibfnamefont {R.}~\bibnamefont {Sensarma}}, \bibinfo {author} {\bibfnamefont {D.}~\bibnamefont {Pekker}}, \bibinfo {author} {\bibfnamefont {E.}~\bibnamefont {Altman}}, \ and\ \bibinfo {author} {\bibfnamefont {E.}~\bibnamefont {Demler}},\ }\href {\doibase 10.1103/PhysRevLett.104.080401} {\bibfield  {journal} {\bibinfo  {journal} {Phys. Rev. Lett.}\ }\textbf {\bibinfo {volume} {104}},\ \bibinfo {pages} {080401} (\bibinfo {year} {2010})}\BibitemShut {NoStop}%
\bibitem [{\citenamefont {Eckstein}\ and\ \citenamefont {Werner}(2011)}]{Eckstein2011}%
  \BibitemOpen
  \bibfield  {author} {\bibinfo {author} {\bibfnamefont {M.}~\bibnamefont {Eckstein}}\ and\ \bibinfo {author} {\bibfnamefont {P.}~\bibnamefont {Werner}},\ }\href {\doibase 10.1103/PhysRevB.84.035122} {\bibfield  {journal} {\bibinfo  {journal} {Phys. Rev. B}\ }\textbf {\bibinfo {volume} {84}},\ \bibinfo {pages} {035122} (\bibinfo {year} {2011})}\BibitemShut {NoStop}%
\bibitem [{\citenamefont {Taranto}\ \emph {et~al.}(2012)\citenamefont {Taranto}, \citenamefont {Sangiovanni}, \citenamefont {Held}, \citenamefont {Capone}, \citenamefont {Georges},\ and\ \citenamefont {Toschi}}]{Taranto2012}%
  \BibitemOpen
  \bibfield  {author} {\bibinfo {author} {\bibfnamefont {C.}~\bibnamefont {Taranto}}, \bibinfo {author} {\bibfnamefont {G.}~\bibnamefont {Sangiovanni}}, \bibinfo {author} {\bibfnamefont {K.}~\bibnamefont {Held}}, \bibinfo {author} {\bibfnamefont {M.}~\bibnamefont {Capone}}, \bibinfo {author} {\bibfnamefont {A.}~\bibnamefont {Georges}}, \ and\ \bibinfo {author} {\bibfnamefont {A.}~\bibnamefont {Toschi}},\ }\href {\doibase 10.1103/PhysRevB.85.085124} {\bibfield  {journal} {\bibinfo  {journal} {Phys. Rev. B}\ }\textbf {\bibinfo {volume} {85}},\ \bibinfo {pages} {085124} (\bibinfo {year} {2012})}\BibitemShut {NoStop}%
\bibitem [{\citenamefont {Werner}\ \emph {et~al.}(2012)\citenamefont {Werner}, \citenamefont {Tsuji},\ and\ \citenamefont {Eckstein}}]{Werner2012}%
  \BibitemOpen
  \bibfield  {author} {\bibinfo {author} {\bibfnamefont {P.}~\bibnamefont {Werner}}, \bibinfo {author} {\bibfnamefont {N.}~\bibnamefont {Tsuji}}, \ and\ \bibinfo {author} {\bibfnamefont {M.}~\bibnamefont {Eckstein}},\ }\href {\doibase 10.1103/PhysRevB.86.205101} {\bibfield  {journal} {\bibinfo  {journal} {Phys. Rev. B}\ }\textbf {\bibinfo {volume} {86}},\ \bibinfo {pages} {205101} (\bibinfo {year} {2012})}\BibitemShut {NoStop}%
\end{thebibliography}%

\end{document}